\documentclass[12pt,dvips]{article}
\usepackage{graphicx}
\begin{document}
\title{Crossover Behaviour of 3-Species Systems with Mutations or 
Migrations} 
\author{Margarita Ifti and Birger Bergersen \\
Department of Physics and Astronomy, University of British Columbia, \\
6224 Agricultural Road, Vancouver, BC, Canada V6T 1Z1}
\maketitle
\normalsize 

\begin{abstract}

We study the $ABC$ model in the cyclic competition ($A + B \rightarrow 
2B$, $B + C \rightarrow 2C$, $C + A \rightarrow 2A$) and the neutral drift 
($A + B \rightarrow 2B$ or $2A$, $B + C \rightarrow 2C$ or $2B$, $C + A 
\rightarrow 2A$ or $2C$) versions, with mutations and migrations 
introduced into the model. When stochastic phenomena are taken into 
account, there are three distinct regimes in the model. $(i)$ In the 
``fixation'' regime, the first extinction time scales with the system size 
$N$ and has an exponential distribution, with an exponent that depends on 
the mutation/migration probability per particle $\mu$. $(ii)$ In the 
``diversity'' regime, the order parameter remains nonzero for very long 
times, and becomes zero only rarely, almost never for large system sizes. 
$(iii)$ In the critical regime, the first passage time for crossing the 
boundary (one of the populations becoming zero) has a power law 
distribution with exponent $-1$. The critical mutation/migration 
probability scales with system size as $N^{-1}$. The transition 
corresponds to a crossover from diffusive behaviour to Gaussian 
fluctuations about a stable solution. The analytical results are checked 
against computer simulations of the model.

\end{abstract}

\section{Introduction}

Cyclic phenomena play a very important role in different classes of 
processes in nature, particularly in epidemiological and ecological 
systems. In the epidemiological context, examples are diseases which do 
not leave us with permanent immunity~\cite{cooke77, longini80}. In 
ecology, cases when three variants of a species compete with one-another 
in a cyclic fashion have been observed~\cite{reeves72, james91, sinervo96, 
smith96}. Another system of interest are cyclic food webs. Recently, Kerr 
et al. obtained laboratory results for a system of three competing species 
that play rock-paper-scissors, when placed in a lattice-like spatial 
structure, and showed that there is agreement of the simulation results to 
the laboratory ones~\cite{kerr02}. Its alternative, as far as evolutionary 
processes are concerned, is the famous Kimura-Weiss model of neutral 
genetic drift~\cite{kimura64, weiss65}. 

In a previous article~\cite{unebirger1} we have considered a non-spatial 
version of the $ABC$ model in both the cyclic competition and neutral 
genetic drift versions, and studied the evolutionary time behaviour of 
such a model. The number of the $A, B, C$ species oscillates with an 
amplitude that drifts with time, until one of the species (and then the 
second one) goes extinct, i.e. biodiversity is lost. The number of 
survivors vs. time plots show an exponential decay. It is very interesting 
to note that there is no difference in the extinction time scale for the 
ensemble of competing species and the case of neutral drift. This allows 
us to think of the neutral model as an ``adiabatic approximation'' of the 
cyclic system. Considering that the cyclic competition system would be a 
minimal model (three alleles) of Darwinian evolution picture, it seems 
that, by just looking at the results, we can not tell which one must have 
been the mechanism for the evolution!

There is growing concern about the effects of habitat fragmentation in the 
survival of a species~\cite{pimm98}. Small, isolated forest fragments lose 
species--we have many examples in the case of fast deforestation 
worldwide~\cite{laura97}. Some species were lost within a few years of 
their isolation from the once-continuous forest~\cite{lovejoy86}. 
Oscillations in the number of population of one ``species'' are observed 
in the case of epidemics, as well as ecology. Throughout the past century 
we have observed changes in patterns of epidemics~\cite{earn00a}. For some 
diseases, the major transitions have been between regular cycles and 
irregular epidemics, and from regionally synchronized oscillations to 
complex, spatially incoherent epidemics. Sinervo et al.~\cite{sinervo96, 
sinervo00} observed spatio-temporal oscillations in the number of lizards 
(male and female) that employ different mating strategies.

Many population ecologists then conclude that a very important issue is 
the synchrony of population dynamics in different habitat 
patches~\cite{brown77, blasius99, earn00, earn98, rohani99}. If the 
population of a certain species goes extinct in one patch while it still 
survives in other patches, then the hope is that what is known as ``rescue 
effect'' can prevent global extinction~\cite{blasius99}. Otherwise, the 
population ecologists say, if there is synchrony of population dynamics 
between patches, the species is doomed to go extinct altogether. In this 
framework, there have been many proposed strategies for preservation, one 
of the most debated of which is the construction of ``conservation 
corridors'' that would make it possible for individuals to move along 
habitat patches~\cite{beier98, gonzalez98}.

Considering that each habitat patch can be thought of as one copy of our 
non-spatial ensemble, it seemed reasonable to study the behaviour of the 
three-species cyclic system in the presence of mutations or migration. In 
the following section we introduce our model, and in the next ones, we 
present the results of our calculations and simulations.

\section{The Model}

Consider a non-spatial (well-stirred) system in which three species $A$, 
$B$, $C$ are competing in a way described by the reactions: $A + B 
\rightarrow 2B$ at a rate $AB/N$, $B + C \rightarrow 2C$, $C + A 
\rightarrow 2A$ at corresponding rates. Add to it the reactions: $A 
\rightarrow B$ at rate $\mu A$, $A \rightarrow C$ at rate $\mu A$, $B 
\rightarrow A$ at rate $\mu B$, $B \rightarrow C$ at rate $\mu B$, $C 
\rightarrow A$ at rate $\mu C$, and $C \rightarrow B$ at rate $\mu C$, 
where $\mu$ is the probability of mutations per individual particle in 
unit time.

The rate equations for this system will be:

\begin{eqnarray}
\frac {dA}{dt} = \frac {AC}{N} - \frac {AB}{N} + \mu B + \mu C -2 \mu A 
\nonumber \\
\frac {dB}{dt} = \frac {BA}{N} - \frac {BC}{N} + \mu A + \mu C -2 \mu B \\
\frac {dC}{dt} = \frac {CB}{N} - \frac {CA}{N} + \mu A + \mu B -2 \mu C 
\nonumber
\end{eqnarray}

\noindent with $A + B + C = N =$ const. If the neutral drift system is 
considered instead, i.e. when $A+B \rightarrow 2A$ or $2B$ with equal 
probability (and similarly for the two other reactions), the rate 
equations will only contain the terms that depend on $\mu$, and the first 
two terms will be absent.

In another version, mutations are replaced by migration into and out of 
the ``island''. In other words, the following migrations are happening: 
$A$ out at rate $3 \mu A$, $B$ out at rate $3 \mu B$, $C$ out at rate $3 
\mu C$, $A$, $B$, or $C$ in at constant rate $\mu \cdot N$, where $N$ is 
the system size (to make it comparable to the rate of leaving the island. 
This would correspond to the average number of individuals in other 
patches, which are the source of our ``immigrants''.) The rate equations 
will be the same as those for the system with mutations. The above 
equations have a fixed point at $A=B=C=N/3$, and it is a stable solution.

However, the rate equations are just a ``mean field'' approximation; they 
only describe the behaviour of the average values of the individual 
populations. In the real world, the system is subject to stochastic noise 
due to birth and death processes (intrinsic noise), which we take to be 
Poisson-distributed. The random nature of these processes need be taken 
into consideration, if we want to get the correct picture of the evolution 
of the system. For that we ought to write the master equation, and then 
try to somehow expand it, obtaining a Fokker-Planck equation, and 
eventually solve it. The two most commonly used expansions of the master 
equation are the Kramers-Moyal expansion, which is essentially a Taylor 
expansion in powers of system size $N$, and the $\Omega$-expansion of van 
Kampen~\cite{vankampen97} which is an expansion in powers of $\sqrt N$. 
The first method produces a nonlinear Fokker-Planck equation. The second 
method is quite systematic, and yields satisfactory results when the 
system has a single stable point, as the rate equations suggest is the 
case with the present model. In the absence of this stable solution, the 
task becomes very difficult~\cite{unebirger1}. 

Using the ``shift'' operators notation:

\begin{eqnarray}
\epsilon_A f(A,B,C) = f(A+1,B,C) \nonumber \\
\epsilon^{-1}_{A} f(A,B,C) = f(A-1,B,C) \nonumber
\end{eqnarray}

\noindent the master equation for the cyclic competition system with 
mutations reads:

\[
\frac{\partial P(A,B,C,t)}{\partial t} = \{ \frac {1}{N} [(\epsilon_C 
\epsilon^{-1}_A -1) AC + (\epsilon_A \epsilon^{-1}_B -1) AB + 
(\epsilon_B \epsilon^{-1}_C -1)BC]+\]\[
+ \mu [(\epsilon_A \epsilon^{-1}_B + \epsilon_A \epsilon^{-1}_C -2)A + 
(\epsilon_B \epsilon^{-1}_C + \epsilon_B \epsilon^{-1}_A -2) B + \]
\begin{equation}\label{eq:mec}
+(\epsilon_C \epsilon^{-1}_A + \epsilon_C \epsilon^{-1}_B  -2) C] \} 
P(A,B,C,t)
\end{equation}

\noindent while that for the neutral drift system with mutations:

\[
\frac{\partial P(A,B,C,t)}{\partial t} = \{ \frac {1}{2N} 
[(\epsilon_C \epsilon^{-1}_A +\epsilon_A \epsilon^{-1}_C -2) AC + \]\[ 
+(\epsilon_A \epsilon^{-1}_B + \epsilon_B \epsilon^{-1}_A -2) AB + 
(\epsilon_B \epsilon^{-1}_C + \epsilon_C \epsilon^{-1}_B -2)BC]+\]\[
+ \mu [(\epsilon_A \epsilon^{-1}_B + \epsilon_A \epsilon^{-1}_C -2)A + 
(\epsilon_B \epsilon^{-1}_C + \epsilon_B \epsilon^{-1}_A -2) B + \]
\begin{equation}\label{eq:nec}
+(\epsilon_C \epsilon^{-1}_A + \epsilon_C \epsilon^{-1}_B  -2) C] \} 
P(A,B,C,t)
\end{equation}

\noindent and that for the cyclic system with migrations:

\[
\frac{\partial P(A,B,C,t)}{\partial t} = \{ \frac {1}{N} [(\epsilon_C 
\epsilon^{-1}_A -1) AC + (\epsilon_A \epsilon^{-1}_B -1) AB + 
(\epsilon_B \epsilon^{-1}_C -1)BC]+\]\[
+ 3 \mu [(\epsilon_A -1)A + (\epsilon_B -1) B + (\epsilon_C -1) C +\]
\begin{equation}\label{eq:mig}
+\frac{N}{3}(\epsilon^{-1}_A + \epsilon^{-1}_B + \epsilon^{-1}_C -3)] \} 
P(A,B,C,t)
\end{equation}

\section{The ``Fixation'' Regime}

In a previous work~\cite{unebirger1} we studied the cyclic competition 
and neutral genetic drift systems in absence of mutations. In the 
mean-field approximation (rate equations) the cyclic system has an 
infinity of neutrally stable solutions: any trajectory that conserves the 
integral $H=ABC/N^3$ is a stable trajectory of the system. On the other 
hand, the neutral drift system has an infinity of neutrally stable points: 
any state of the system is stable in the mean-field picture. The story is 
different, when stochastic birth-and-death processes are taken into 
account: then the individual population distribution drifts, until one of 
the species, and then another one, go extinct. In other words, there is 
fixation of the population to one of the varieties, and the product 
$H=ABC/N^3$ becomes zero. The survival probability versus time (in units 
of $N$) plot exhibits exponential decay with slope $-3$, which was 
verified by numerically solving the corresponding Fokker-Planck equation. 
This equation was obtained from a Kramers-Moyal expansion, since the 
infinite multiplicity of stable solutions (points or limit cycles) makes 
the van Kampen method impractical to apply. The first extinction time 
scales with the total population size $N$. This behaviour is the same for 
both the cyclic competition and the neutral drift models, the second one 
behaving as an ``adiabatic approximation'' to the first one. For both 
models, the probability distribution function becomes uniform within a 
short time, i.e. any point inside the triangle that represents the phase 
space of the system is equally probable.

When $\mu$ is small (the meaning of small will become clear in section 5) 
the system with mutations/migrations is in the ``fixation'' regime, in 
which the results of our previous study~\cite{unebirger1} are applicable 
qualitatively. In that regime, the stochastic processes push the system 
towards the (absorbing) boundary, in which at least one of the species has 
met extinction. This was verified by computer simulations of the master 
equations~(\ref{eq:mec}, \ref{eq:nec}, \ref{eq:mig}). These simulations 
started with equal individual populations of $A$, $B$, $C$, (i.e. the 
centre). We generated times for the next possible reaction event with 
exponential distribution as $- \frac {\ln (rn)}{rate}$, where the rate of 
the cyclic/neutral or mutations/migrations as in the master equations 
above~(\ref{eq:mec}, \ref{eq:nec}, \ref{eq:mig}) is substituted. (Here 
$rn$ is a random variable with uniform distribution in $[0,1]$. This way 
we get Poisson distribution for the event times, i.e. really independent 
events~\cite{gibson}.) The reaction which occurs first is then picked and 
the system is updated accordingly. The process is repeated for a large 
number of events.

\begin{figure}
\includegraphics{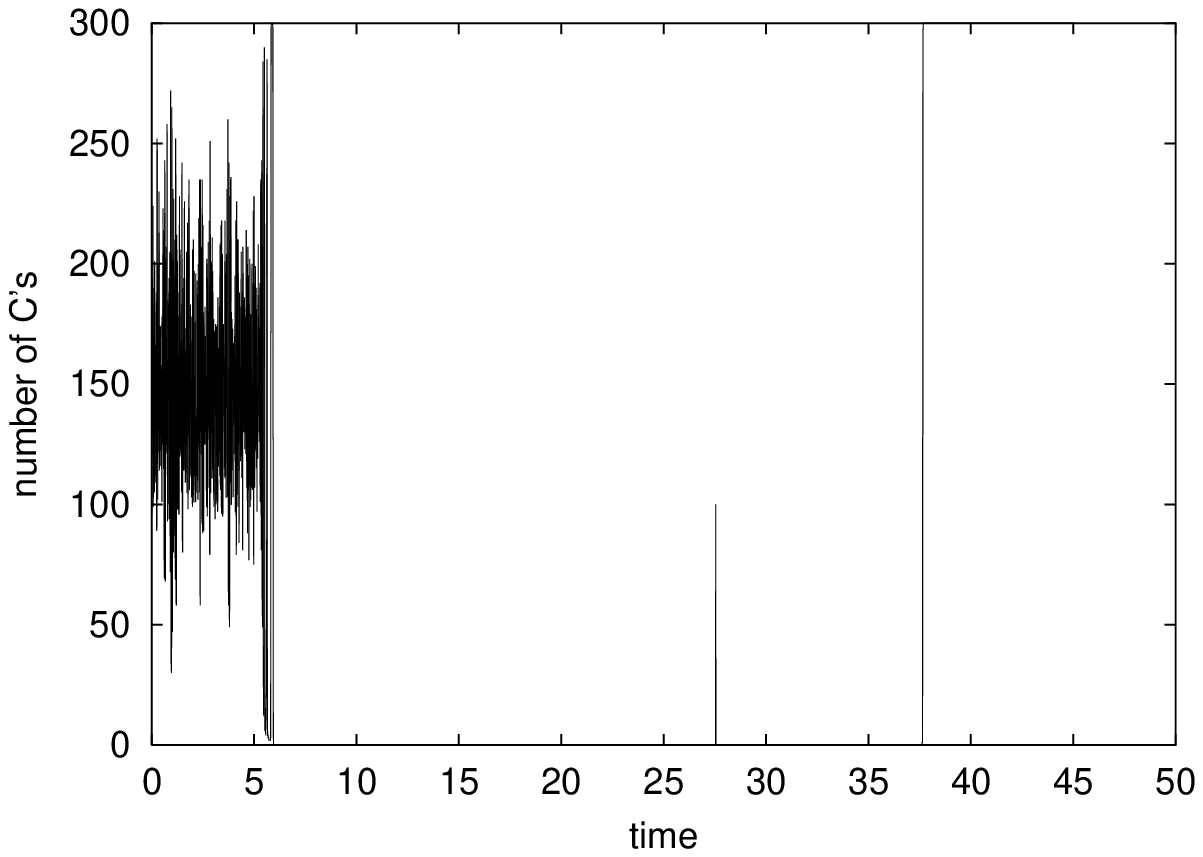}
\caption{The time series for the number of $C$ species in the 
``fixation'' regime (here $\mu=0.4 \cdot 10^{-3}$, and system size 
$N=300$).}
\label{fig:tsext}
\end{figure}

We observed a ``fixation'' regime, in which one of the populations, and 
then the next one, go to zero quite fast. In this regime, the number of 
individual populations oscillates with an amplitude that drifts with time. 
The main reaction is then the cyclic one, and there are only occasional 
mutations/migrations, which are not frequent enough to prevent fixation. 
In other words, the system is experiencing ``forces'' due to fluctuations, 
and ``forces'' due to mutations (migrations). The fluctuations are pushing 
the system toward the boundary, while the mutations (migrations) push 
toward the centre. For small mutation/migration rate the 
``fluctuation-generated forces'' are stronger. Fig.~\ref{fig:tsext} shows 
the variation with time of the population of $C$'s for a realisation of 
the system in the ``extinction'' regime. Occasionally, an individual of 
the competing species (the one next in the ``food chain'') is introduced 
in the system by mutations (or migrations), and then it either causes an 
occasional spike in the time series (like the one we see in 
Fig.~\ref{fig:tsext} at $25<t<30$, or it ``eats up'' the old species 
completely and replaces it in the system (which is what happens between 
$t=35$ and $t=40$ in our plot).

\begin{figure}
\includegraphics{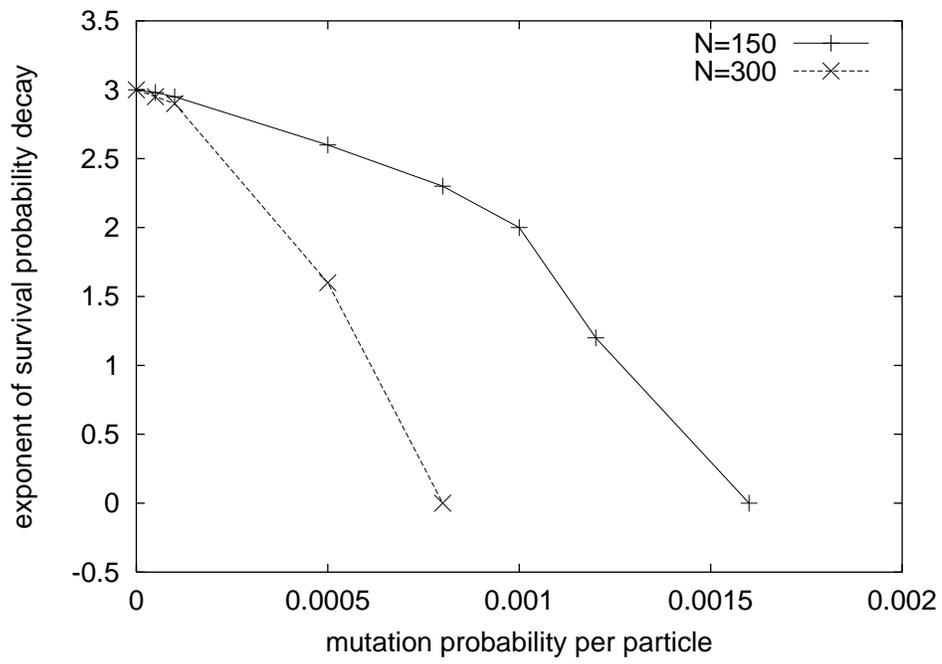}
\caption{Dependence of the exponent of the decay of the survival 
probability on the mutation rate in the ``extinction'' regime for system 
size $N=150$.}
\label{fig:expvsmu}
\end{figure}

We investigated the behaviour of the probability distribution for the 
first crossing of the boundary, i.e. when one of the populations becomes 
zero for the first time, as the mutation/migration rate increases. For 
that we ran ten thousand copies of the system for each value of the 
mutation rate. The survival probability (as defined above) was plotted vs. 
time. In the ``fixation regime'' the decay is still exponential, but the 
exponent varies with the mutation/migration rate, getting close to zero as 
the system approaches criticality. Fig.~\ref{fig:expvsmu} shows a plot of 
the exponent as a function of mutation probability $\mu$ for system size 
$N=150$ and $N=300$.

\section{The ``Diversity'' Regime}

If only mutations (migrations) are present, the system remains near the 
centre point ($A=B=C=N/3$), and the order parameter remains considerably 
above zero; in other words, all three species are present in the system. 
One can occasionally observe temporary extinctions, but this happens very 
seldom, and for very long times; when the system size is very large, it 
almost never happens. Since the boundary is not absorbing, occasional 
mutations/migrations will return the system to the state with maximal 
symmetry (biodiversity) where all three varieties coexist. When both the 
cyclic/neutral drift mechanism and mutations (migrations) are present, and 
the $\mu$ is above critical, the mutations manage to keep the system 
maximally disordered, since they are stronger than the fluctuations 
(which, as we saw, try to drive the system toward the boundary, i.e. 
fixation, and keep it there). Compared to the situation with no/low 
mutations (migrations), where the boundary is (practically) absorbing, and 
the final state of the system is a ``pure'' one (with only one of the 
species present), the mutation/migration rate acts then as a 
``temperature''.

Work has been done on the two allele almost neutral drift model with 
mutations~\cite{tim}. The almost neutral model with mutations, preserving 
the total number of individuals, has only one degree of freedom, and 
allows one to derive an ``effective potential'' from the Fokker-Planck 
equation, obtained by a Kramers-Moyal expansion of the master equation. 
For small mutation probabilities, such that $2\mu N \ll 1$, there is 
extinction of one species and fixation. The effective potential is 
symmetric around the centre (where both species are in equal numbers) and 
the branches of the effective potential are down. This allows for the 
system to quickly slip into the state where only one of the species is 
present. Otherwise, both species coexist forever in the high mutation 
regime, i.e. when $2 \mu N \gg 1$. In that regime, the effective potential 
is symmetric around the centre point, but with branches upwards, which 
means that the centre point is a minimum potential point. The system then 
remains in the vicinity of that point for very long times. The effective 
potential ``flips'' from ``branches up'' to ``branches down'' at the point 
where $2 \mu N =1$. The transition is second-order, and critical behaviour 
is observed. The system behaves similarly when migrations are present, 
instead of mutations. One aspect of migrations in a four-species system 
has been treated recently by Togashi and Kaneko~\cite{togashi01}.

Since the rate equations have a stable solution, we employ the van Kampen 
expansion~\cite{vankampen97}. The idea of this expansion is to split the 
variables of the problem into a non-fluctuating part, and a fluctuating 
one, i.e. deal separately with the mean-field solutions and the 
fluctuations. In this approach, the numbers of the individual populations 
would be written as:

\begin{eqnarray}
A=N \phi_1 + \sqrt N x_1 \nonumber \\
B=N \phi_2 + \sqrt N x_2 \\
C=N \phi_3 + \sqrt N x_3 \nonumber
\end{eqnarray}

\noindent Here the $\phi_i$ are the concentrations of $A$, $B$, and $C$ 
species respectively (which only depend on time), and the $x_i$ are the 
fluctuations (proportional to the square root of system size). The system 
size (total population) $N$ is conserved for the system with mutations, 
but not for that with migrations. This conservation rule will cause 
trouble in the case of the system with mutations, and will require special 
attention. Using the van Kampen Ansatz, the probability distribution 
$P(A,B,C,t)$ is transformed into $\Pi(\{x_i\},t)$, and the following 
relations are true:

\begin{eqnarray}
\Pi = N^{3/2} P(N \{\phi_i + \sqrt N x_i\},t) \nonumber \\
\frac {\partial P}{\partial t} = \frac {1}{N^{3/2}} \frac {\partial 
\Pi}{\partial t} - \frac{1}{N} \sum \frac {d \phi_i}{d t} \frac 
{\partial \Pi}{\partial x_i} \nonumber
\end{eqnarray}

\noindent and

\begin{eqnarray}
\epsilon_i = 1+ \frac{1}{\sqrt N} \frac {\partial}{\partial x_i} + \frac 
{1}{2N} \frac {\partial^2}{\partial {x_i}^2} + \ldots \\
{\epsilon_i}^{-1} = 1- \frac{1}{\sqrt N} \frac {\partial}{\partial x_i} + 
\frac {1}{2N} \frac {\partial^2}{\partial {x_i}^2} + \ldots \nonumber
\end{eqnarray}

Next we substitute everything into the master equation, leave only the 
term $\partial \Pi / \partial t$ in the left hand side, and group the 
right hand side terms according to powers of $\sqrt N$. The first term is 
of order $N^{1/2}$, and it must be equal to zero, for an expansion in 
terms of $N^{1/2}$ to make sense. That term for e.g. the cyclic model is:

\begin{equation}
\sum \frac {\partial \Pi}{\partial x_i} [\frac {d \phi_i}{dt} + \phi_i 
\phi_{i+1} - \phi_i \phi_{i+2} + \mu (2 \phi_1 - \phi_{i+1} - 
\phi_{i+2})] = 0,
\end{equation}

\noindent which reproduces the rate equations in terms of the 
concentrations $\phi_i$, with steady state solution $\phi_i=1/3$. 
Similarly, we get the rate equations for the other models.

The terms of order $N^0$ give a linear Fokker-Planck equation of the 
form:

\begin{equation}\label{eq:fpsurv}
\frac {\partial \Pi}{\partial t} =\sum [-A_{ik} \frac {\partial}{\partial 
x_i} (x_k \Pi) + \frac{1}{2} B_{ik} \frac {\partial^2 \Pi}{\partial x_i 
\partial x_k}]
\end{equation}

\subsection{The System with Migrations}

We are going to solve the system with migrations first, since the absence 
of the conservation of total population $N$ makes this system easier to 
deal with. For simplicity, let us limit our attention to fluctuations 
around the steady state $\phi_i=1/3$. The A-matrix for the system with 
migrations is:

\[ \left(
\begin{array}{ccc}\label{eq:amatmig}
- 3 \mu & - 1/3 & 1/3 \\
1/3 & -3 \mu & - 1/3 \\
- 1/3 & 1/3 & -3 \mu \end{array}
\right) \]

\noindent and the B-matrix:

\[ \left(
\begin{array}{ccc}\label{eq:bmatmig}
2 \mu+2/9 & -1/9 & -1/9 \\
-1/9 & 2\mu+2/9 & -1/9 \\
-1/9 & -1/9 & 2\mu+2/9 \end{array}
\right) \]

These equations are linear, and the coefficients depend on time through 
$\phi_i$. This approximation is otherwise known as ``linear noise 
approximation''. The solution is known to be a Gaussian; the problem 
represents itself as an Ornstein-Uhlenbeck process. For our purposes, it 
suffices to determine the first and second moments of the fluctuations. 
Following van Kampen~\cite{vankampen97}, we can multiply the Fokker-Planck 
equation by $x_i$ and integrate by parts to get:

\begin{equation}
\frac {d <x_i>}{dt} = \sum_j A_{ij} <x_j>
\end{equation}

\noindent The eigenvalues of the $A$ matrix are of the form $-3 \mu$, 
$-3 \mu \pm i/ \sqrt 3$. The negativity of the eigenvalues guarantees the 
stability of the zero solutions to the first moments equations. Hence, the 
average of the fluctuations decays to zero and remains zero. The equations 
for the second moments can be obtained similarly:

\begin{equation}\label{eq:2ndmom}
\frac {d <x_i x_j>}{dt} = \sum_k A_{ik} <x_k x_j> + \sum_k A_{jk} <x_i 
x_k> + B_{ij}
\end{equation}

By symmetry, in the steady state all the diagonal terms $<x_i^2>$ are 
equal, as well as off-diagonal terms (correlations) $<x_i x_j>$. They 
depend on the migration probability $\mu$ alone. The steady state 
solutions are:

\begin{eqnarray}\label{eq:solmig}
<x_i^2> = \frac {9 \mu + 1}{27 \mu} \\
<x_i x_j> = - \frac{1}{54 \mu} \nonumber
\end{eqnarray}

\begin{figure}
\includegraphics{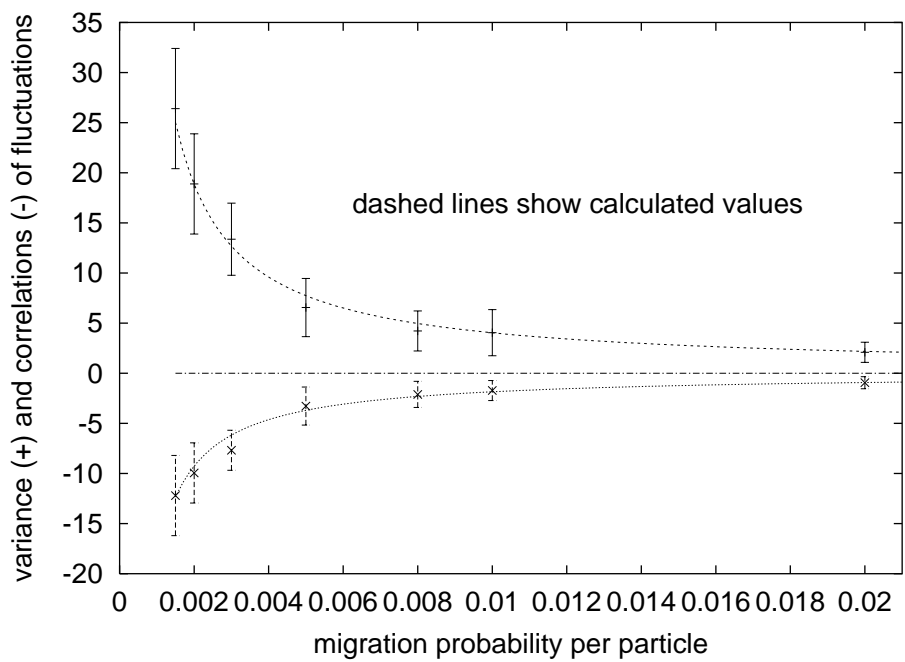}
\caption{The variance and correlations of the fluctuations vs. migration 
probability per particle for system size $N=300$, cyclic system. The same 
behaviour is observed for the system with mutations.}
\label{fig:flucmig}
\end{figure}

\noindent The diagonal terms coincide with the variance, since the 
$<x_i>=0$. These fluctuations were measured ``experimentally'', i.e. 
calculated from the results of the simulations. We simulated 1000 copies 
of the system (size $N=300$) at different mutation rates, and calculated 
the mean and variance of the fluctuations, as well as correlations. The 
results of those simulations are shown in Fig.~\ref{fig:flucmig} for the 
system with migrations. They agree with the calculated values.

\subsection{The System with Mutations}

In the case of the system with mutations, the total population $N$ is a 
conserved quantity; this imposes restrictions on allowed fluctuations. We 
avoid this problem by excluding one of the variables. For that, we 
transform the problem into one with two variables, which can be done by 
putting:

\begin{eqnarray}
\frac{A}{N}=\frac{1}{3} - \frac{1}{\sqrt N} (\frac{\sqrt 3}{2} x 
+\frac{y}{2}) \nonumber \\
\frac{B}{N}=\frac{1}{3} + \frac{1}{\sqrt N} (\frac {\sqrt 3}{2} x - \frac 
{y}{2}) \\
\frac{C}{N}=\frac{1}{3} + \frac{y}{\sqrt N} \nonumber
\end{eqnarray}

This corresponds to transforming to Cartesian coordinates with the origin 
placed at the geometrical centre of the equilateral triangle, which 
constitutes the phase space of our system~\cite{unebirger1}.

The Fokker-Planck equation for the cyclic competition system becomes:

\begin{equation}
\frac{\partial \Pi}{\partial t} = \frac {\partial}{\partial x} (\frac{y} 
{\sqrt 3} + 3\mu x) \Pi - \frac{\partial}{\partial y} (\frac{x}{\sqrt 3} - 
3\mu y) \Pi + \frac{1+6\mu}{9}(\frac{\partial^2}{\partial x^2}+ 
\frac{\partial^2}{\partial y^2}) \Pi
\end{equation}

\noindent The A-matrix has the form:

\[ \left(
\begin{array}{cc}
- 3 \mu & - 1/ \sqrt 3 \\
1/ \sqrt 3 & -3 \mu \end{array}
\right) \]

\noindent with eigenvalues $-3 \mu \pm i/ \sqrt 3$. The mean value of the 
fluctuations decays to zero with an oscillating behaviour, and remains 
zero. The B-matrix is diagonal, with both diagonal elements equal to 
$\frac{2}{9}(1+6\mu)$. The equations for the second moments become:

\begin{eqnarray}
\frac{d<x^2>}{dt} = -6\mu<x^2> - \frac{2}{\sqrt 3} <xy> + 
\frac{2}{9}(1+6\mu) \nonumber \\
\frac{d<xy>}{dt} = \frac{2}{\sqrt 3}<x^2> -6\mu<xy> - \frac{2}{\sqrt 3} 
<y^2> \\
\frac{d<y^2>}{dt} = \frac{2}{\sqrt 3} <xy> -6\mu<y^2> + 
\frac{2}{9}(1+6\mu) \nonumber
\end{eqnarray}

\noindent with solutions that tend to

\begin{equation}
<x^2>=<y^2>=\frac{1+6\mu}{27\mu}, \ \ <xy>=0
\end{equation}

\noindent as $t \rightarrow \infty$. These solutions need to be 
transformed back in terms of $<x_i^2>$ and $<x_i x_j>$, and the results 
are:

\begin{eqnarray}
<x_i^2> = \frac{1+6\mu}{27\mu} \\
<x_i x_j> = -\frac{1+6\mu}{54\mu} \nonumber
\end{eqnarray}

For the neutral drift case, the Fokker-Planck equations becomes:

\begin{equation}
\frac{\partial \Pi}{\partial t} = \frac {\partial}{\partial x}(6\mu x 
\Pi) + \frac{\partial}{\partial y} (6\mu y \Pi) + \frac{2(1+6\mu)} {9} 
(\frac{\partial^2}{\partial x^2}+ \frac{\partial^2}{\partial y^2}) \Pi
\end{equation}

\noindent with the same steady state distributions as the cyclic 
competition model.

\section{The Transition Region}

The qualitative change in the behaviour of the system, when the value of 
the parameter $\mu$ is varied, speaks of the presence of a phase 
transition. The survival probability decays exponentially with time in the 
``fixation'' regime, and the exponent goes down as the mutation 
probability increases, as shown in Fig.~\ref{fig:expvsmu}. This speaks of 
a symmetry-breaking transition as the mutation probability goes through 
the critical value.

\begin{figure}
\includegraphics{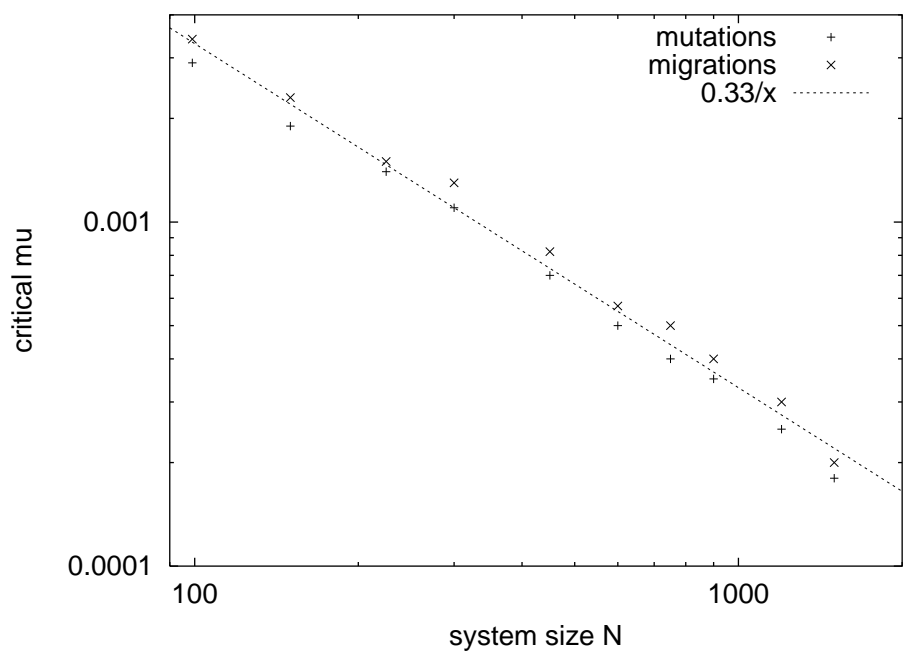}
\caption{Variation of critical mutation and migration probability per 
particle with system size.}
\label{fig:mucrit}
\end{figure}

When the mutation/migration probabilities per particle approach zero, the 
leading term in the variances of the concentrations of individual 
populations is of order $1/27 N \mu$, and it becomes of the same order of 
magnitude as the macroscopic concentrations, when $3 \mu N \propto 1$. 
This gives us the critical mutation/migration probability dependence on 
the system size $N$. The critical $\mu$ was verified to be the same for 
both models with mutations present, and the critical $\mu$ for the system 
with migrations from the simulations is very close to that for the systems 
with mutations, which supports the calculations above. 
Fig.~\ref{fig:mucrit} shows plots of critical $\mu$ vs. $N$, as well as a 
line of slope $1/N$ for comparison. To further verify this result, we 
performed a Kramers-Moyal expansion the models with mutations and 
migrations, in which $\mu$ was posed as $\mu_0/N$. Numerical solution of 
the resulting nonlinear Fokker-Planck equations gave the equation for the 
eigenvalue (exponent of decay): $\lambda = 9 \mu_0 -3$, where $\lambda$ is 
the eigenvalue. At criticality, $\lambda=0$, which yields $\mu_0=1/3$ or 
$\mu=1/3N$~\cite{thesis}. This is in excellent agreement with the results 
obtained from the analysis above, as well as the computer simulations. It 
is worth noting that the neutral system essentially behaves like the 
cyclic one.

The probability that the system has never crossed the boundary, (i.e. none 
of the populations has ever become zero,) when the system is in the 
transition regime, still decays to zero, but not exponentially any more. 
The transition is critical; however, transitions in nonequilibrium 
(steady-state) systems are different from the thermodynamic phase 
transitions. If the system size were infinite, one would introduce one 
infected individual (mutant) and measure the probability of survival of 
infection/mutation as time goes to infinity. This would constitute the 
order parameter of our system in the thermodynamic limit. However, our 
systems are finite, and for finite $N$ the quantity which exhibits 
critical behaviour is the first passage time for crossing the 
boundary~\footnote{Our model differs from the chemical reactions models: 
in them the onset of the cyclic behaviour is a Hopf bifurcation, in which 
a stable focus changes into a limit cycle~\cite{jackson}. In our model the 
limit cycle is absent. The chemical reactions models are dissipative even 
in the mean-field approximation, while our system has a centre in the mean 
field treatment. In those models the fluctuations become important only 
when the system is in the vicinity of the Hopf bifurcation, ours is 
entirely fluctuations-driven below the critical transition point.}.

\begin{figure}
\includegraphics{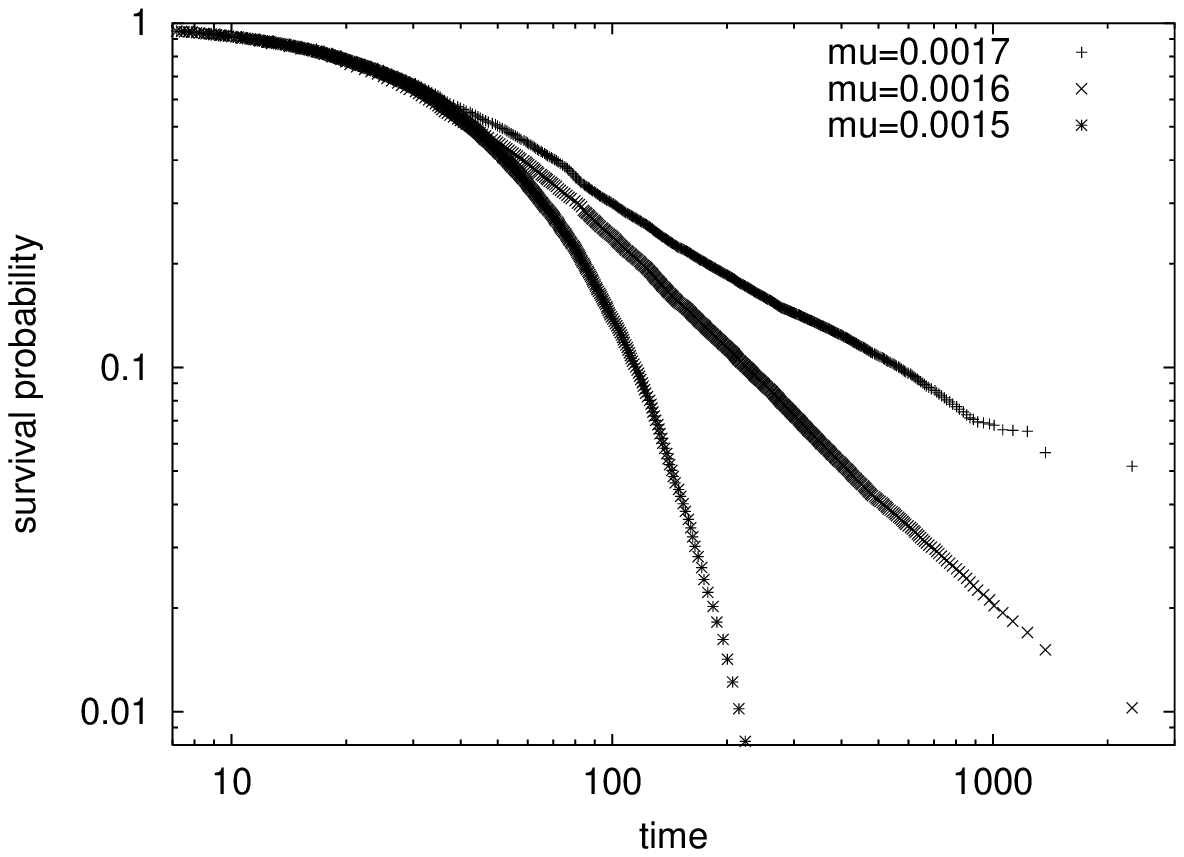}
\caption{Cummulative survival probability vs. time (in units of $N$) just 
below, just above, and at the critical point, cyclic system with 
mutations, system size $N=210$.}
\label{fig:plcrit}
\end{figure}

In Fig.~\ref{fig:plcrit} we show plots of the first extinction times (with 
their cummulative probability in the y-axis) just below, just above, and 
at the critical $\mu$ (system size $N=210$). (It is worth mentioning that 
we ran a statistics of 10000 copies of the system, but we are showing only 
1 in 20 points in that plot, to prevent figures from becoming cluttered.) 
The power-law decay behaviour was used as a criterion for determining the  
critical point. The plot in Fig.~\ref{fig:plcrit} shows clearly that the 
survival probability decays as a power law at the critical point. However, 
once the system is above the critical point, there is considerable 
probability for the system to have never ``gone dead'', even at very long 
times. The extinction times at the critical point scale with the system 
size. The power-law exponent is $-1.03 \pm 0.04$ for cyclic competition 
model, $-0.99 \pm 0.03$ for the neutral model with mutations; and $-1.05 
\pm 0.06$ for the cyclic system with migrations. This exponent close to 
1, found in simulations is compatible with the one expected from branching 
processes~\cite{resnick94}, as well as the one obtained for the 
two-species Kimura-Weiss model~\cite{tim}.

\section{Conclusions}

We have considered an $ABC$ model with cyclic competition/neutral drift 
and mutations (migrations) at a constant probability. The system 
exhibits a critical transition from a ``fixation'' regime to one in 
which biodiversity is preserved over long time. In the ``fixation'' 
regime, the number of the $A, B, C$ species oscillates with an 
amplitude that drifts with time, until one of the species (and then the 
second one) goes extinct, i.e. the order parameter, defined as the 
product of $ABC/N^3$ goes to zero, and remains zero, except for 
occasional ``bursts''. In the ``diversity'' regime, the number of the 
$A, B, C$ varieties fluctuates around the centre point, and there are 
rare extinctions, but the order parameter remains nonzero almost 
always. The survival probability decays exponentially below the 
transition point, but the exponent decreases as the mutation 
(migration) probability per particle increases, until it becomes zero 
at the critical point. The critical mutation/migration probability 
depends on system size as $N^{-1}$, and the models have the same 
power-law exponent: -1. There is no difference in the behaviours of the 
neutral system and the cyclic system. Also, there is no qualitative 
difference between the system with mutations and that with migrations.

These results address the concern about the effects of habitat 
fragmentation in the survival of the species~\cite{pimm98}. If the 
population of a certain species goes extinct in one patch (e.g. a herd, 
school, swarm) while it still survives in other patches, then it is hoped 
that the ``rescue effect'' can prevent global extinction~\cite{blasius99, 
brown77, earn00}. Otherwise, the species is doomed to go extinct 
altogether. Our results suggest that the number of mutants/migrants 
necessary to preserve diversity is independent of system size, while in 
most realistic situations the number of mutations that happen would be 
proportional to system size. On the other hand, if one has in mind 
epidemiological systems, the important factor is to reduce migration 
probabilities below critical, which is exactly the purpose quarantines 
serve. Let's recall here the revival of SARS epidemics in Toronto, once 
the guard was let down, i.e. infected individuals were allowed to migrate 
from one community on to others. In that case, a large system size may be 
a disadvantage. It takes only one bad apple...

In lattice models~\cite{rosen96, schafer98} small amounts of migration
bring about ``phase synchronization''. Peak population abundances, 
however, are observed to be largely uncorrelated. Blasius et 
al.~\cite{blasius99} use a Langevin type system of equations to introduce 
noise in the three-species spatially structured model, and obtain complex 
chaotic travelling-wave structures.

This work on three-species ecological or epidemiological systems 
relates to autocatalytic systems with a loop-like 
structure~\cite{jain02, togashi01}. Methods of analysis employed in 
this paper can be expanded to the loop-like autocatalytic systems. We 
have work in progress regarding the four- or more-species autocatalytic 
loops~\cite{unebirger3}.

\begin{thebibliography}{99}

\bibitem{cooke77} Cooke~K.~L., Calef~D.~F., and Level~E.~V., {\it 
Nonlinear Systems and its Applications}, Academic Press, New York, 
73 (1977).

\bibitem{longini80} Longini~I.~M., Mathematical Biosciences, 
${\bf 50}$, 85 (1980).

\bibitem{reeves72} Reeves~P., {\it The Bacteriocins}, Springer Verlag, 
New York (1972).

\bibitem{james91} James~R., Lazdunski~C., and Pattus~F., (editors) 
{\it Bacteriocins, Microcins and Lantibiotics}, Springer Verlag, New York 
(1991).

\bibitem{sinervo96} Sinervo~B. and Lively~C., Nature, ${\bf 380}$, 
240 (1996).

\bibitem{smith96} Maynard Smith~J., Nature, ${\bf 380}$, 198 (1996).

\bibitem{kerr02} Kerr~B., Riley~M.~A., Feldman~M.~W., and 
Bohannan~B.~J.~M., Nature, {\bf 418}, 171 (2002).

\bibitem{kimura64} Kimura~M., and Weiss~G.~H., Genetics, ${\bf49}$, 
561 (1964).

\bibitem{weiss65} Weiss~G.~H., and Kimura~M., J. Appl. Prob., ${\bf2}$, 
129 (1965).

\bibitem{unebirger1} Ifti~M., and Bergersen~B., Eur. Phys. J. E, {\bf 
10(3)}, 241 (2003).

\bibitem{pimm98} Pimm~S.~L., Nature, ${\bf 393}$, 23 (1998).

\bibitem{laura97} Laurance~W.~F., and Bierregaard~R.~O. (eds) 
{\it Tropical Forest Remnants: Ecology, Management, and Conservation of 
Fragmented Communities}, Univ. Chicago Press (1997).

\bibitem{lovejoy86} Lovejoy~T.~E. et al. in {\it Conservation Biology: 
The Science of Scarcity and Diversity} (ed. Soul\'e~M.~E.), Sinauer, 
Sunderland, MA (1986).

\bibitem{earn00a} Earn~D.~J.~D., Rohani~P., Bolker~B.~M., and 
Grenfell~B.~T., Science, ${\bf 287}$, 667 (2000).

\bibitem{sinervo00} Sinervo~B, Svensson~E., and Comendant~T., Nature, 
${\bf 406}$, 985 (2000).

\bibitem{brown77} Brown~J.~H., and Kodric-Brown~A., Ecology, 
${\bf 58}$, 445 (1977).

\bibitem{blasius99} Blasius~B., Huppert~A., and Stone~L., Nature, 
${\bf 399}$, 354 (1999).

\bibitem{earn00} Earn~D.~J.~D., Levin~S.~A., and Rohani~P., Science, 
${\bf 290}$, 1360 (2000).

\bibitem{earn98} Earn~D.~J.~D., Rohani~P., and Grenfell~B.~T., Proc. R. 
Soc. London Ser. B, ${\bf 265}$, 7 (1998).

\bibitem{rohani99} Rohani~P., Earn~D.~J.~D., and Grenfell~B.~T., 
Science, ${\bf 286}$, 968 (1999).

\bibitem{beier98} Beier~P., and Noss~R.~F., Conserv. Biol., ${\bf 12}$, 
1241 (1998).

\bibitem{gonzalez98} Gonzalez~A., Lawton~J.~H., Gilbert~F.~S., 
Blackburn~T.~M., and Evans Freke~I., Science ${\bf 281}$, 2045 (1998).

\bibitem{vankampen97} van Kampen~N.~G., {\it Stochastic Processes in 
Physics and Chemistry}, North Holland (1997).

\bibitem{gibson} Gibson~M.~A., and Bruck~J., J. Phys. Chem. A, 
{\bf 104}, 1876 (2000).

\bibitem{tim} Duty~T., Ph.D. thesis, University of British Columbia 
(2000).

\bibitem{togashi01} Togashi~Y., and Kaneko~K., Phys. Rev. Let., 
{\bf 86 (11)}, 2459 (2001).

\bibitem{thesis} Ifti~M., Ph.D. thesis, 
www.physics.ubc.ca/\textasciitilde ita/thesis.pdf.

\bibitem{jackson} Jackson~E.~A., {\it Perspectives of nonlinear dynamics}, 
Cambridge University Press (1991).

\bibitem{resnick94} Resnick~S., {\it Adventures in Stochastic Processes}, 
Springer-Verlag, New York (1994).

\bibitem{rosen96} Rosenblum~M.~G., Pikovsky~A.~S., and Kurths~J., Phys. 
Rev. Lett., ${\bf 76}$, 1804 (1996).

\bibitem{schafer98} Schafer~C., Rosenblum~M.~G., Kurths~J., and 
Abel~H., Nature, ${\bf 392}$, 239 (1998).

\bibitem{jain02} Jain~S., and Krishna~S., arXiv:nlin.AO/0210070 (30 Oct 
2002).

\bibitem{unebirger3} Ifti~M., and Bergersen~B., to be published.

\end{thebibliography}
\end{document}